\documentclass[12pt,preprint]{aastex}
\usepackage{natbib}
\usepackage{url}
\usepackage{amsmath}
\usepackage{lscape}
\usepackage{appendix}
\usepackage{hyperref}
\usepackage{geometry}

\begin{document}

\title{Spatial segregation of massive clusters in dwarf galaxies}

\author{Bruce G. Elmegreen\altaffilmark{1},
A. Adamo\altaffilmark{2},
M. Boquien\altaffilmark{3},
F. Bournaud\altaffilmark{4},
D. Calzetti\altaffilmark{5},
D. O. Cook\altaffilmark{6},
D.A. Dale\altaffilmark{7},
P.-A. Duc\altaffilmark{4,8},
D. M. Elmegreen\altaffilmark{9},
J. Fensch\altaffilmark{10},
K. Grasha,\altaffilmark{11,12},
Hwi Kim\altaffilmark{13},
L. Kahre\altaffilmark{14},
M. Messa\altaffilmark{5},
J. E. Ryon\altaffilmark{15},
E. Sabbi\altaffilmark{15},
L.J. Smith\altaffilmark{16}}
\altaffiltext{1}{IBM Research Division, T.J. Watson Research Center, Yorktown Hts., NY 10598}
\altaffiltext{2}{Dept. of Astronomy, The Oskar Klein Centre, Stockholm University, Stockholm, Sweden}
\altaffiltext{3}{Centro de Astronomía (CITEVA), Universidad de Antofagasta, Avenida Angamos 601, Antofagasta, Chile}
\altaffiltext{4}{AIM, CEA, CNRS, Universit\'e Paris-Saclay, Universit\'e Paris Diderot, Sorbonne Paris Cit\'e, 91191 Gif-sur-Yvette, France}
\altaffiltext{5}{Dept. of Astronomy, University of Massachusetts Amherst, Amherst, MA 01003, USA}
\altaffiltext{6}{California Institute of Technology, Pasadena, CA}
\altaffiltext{7}{Dept. of Physics and Astronomy, University of Wyoming, Laramie, WY, USA}
\altaffiltext{8}{Universit\'e de Strasbourg, CNRS, Observatoire astronomique de Strasbourg, UMR 7550, 67000 Strasbourg, France}
\altaffiltext{9}{Vassar College, Dept. of Physics and Astronomy, Poughkeepsie, NY, USA}
\altaffiltext{10}{Univ. Lyon, ENS de Lyon, Univ. Lyon 1, CNRS, Centre de Recherche Astrophysique de Lyon, UMR5574, F-69007 Lyon, France}
\altaffiltext{11}{Research School of Astronomy and Astrophysics, Australian National University, Canberra, ACT 2611, Australia}
\altaffiltext{12}{ARC Centre of Excellence for All Sky Astrophysics in 3 Dimensions (ASTRO 3D), Australia}
\altaffiltext{13}{NSF's National Optical-Infrared Astronomy Research Laboratory/Gemini Observatory, Casilla 603, La Serena, Chile}
\altaffiltext{14}{Department of Astronomy, New Mexico State University, Las Cruces, NM, USA}
\altaffiltext{15}{Space Telescope Science Institute, Baltimore, MD 21218, USA}
\altaffiltext{16}{European Space Agency/Space Telescope Science Institute, Baltimore, MD, USA}

\begin{abstract}
The {\it relative average minimum projected separations} of star clusters in the
Legacy ExtraGalactic UV Survey (LEGUS) and in tidal dwarfs around the interacting
galaxy NGC 5291 are determined as a function of cluster mass to look for
cluster-cluster mass segregation. Class 2 and 3 LEGUS clusters, which have a more
irregular internal structure than the compact and symmetric class 1 clusters, are
found to be mass segregated in low mass galaxies, which means that the more
massive clusters are systematically bunched together compared to the lower mass
clusters. This mass segregation is not present in high-mass galaxies nor for class
1 clusters. We consider possible causes for this segregation including differences
in cluster formation and scattering in the shallow gravitational potentials of low
mass galaxies.
\end{abstract}

\section{Introduction}
Compact star clusters usually form inside more extended associations of young stars
\citep{feitzinger87,elmegreen97,hunter99,maiz01,lada03,elmegreen06,elmegreen08,portegies10,krumholz19},
as part of a hierarchical structure for star formation that resembles the
distribution of dense interstellar clouds
\citep{scalo85,fleck96,elmegreen96,cartwright04}.  The relative positions and ages
of these clusters follow power-law correlations \citep{efremov98,fuente09,grasha17a}
suggesting this hierarchy is the result of turbulent motions with self-gravity
dominating the densest phase.

While this basic structure is well observed, there has been little effort to
quantify the spatial correlation as a function of cluster mass. We do not know, for
example, if the most massive clusters group together with lower mass clusters
surrounding them. Such mass segregation can be an important constraint on cluster
formation models and an indicator of the history of the region, including
competitive \citep[e.g.,][]{zinnecker82,bonnell97} or cooperative
\citep{elmegreen14} accretion of gas into the clusters, or density-dependent cloud
masses \citep{alfaro18}. Mutual cluster attraction leading to coalescence in dense
regions \citep{lahen19} might also be indicated.

Here we describe a new metric for mass-dependent clustering that has the potential
to reveal whether clusters segregate according to mass. We apply this metric to
clusters in the Legacy ExtraGalactic UV Survey (LEGUS) database \citep{calzetti15}
and, by comparison, to clusters in tidal dwarfs around the interacting galaxy NGC
5291 \citep{fensch19}. These tidal dwarfs have HST images and the highest number of
clusters observed so far for this type of galaxy.

\section{Method}
\label{method}
\subsection{The Relative Average Minimum Projected Separation}

To gain insight on mass segregation among high and low mass clusters, we consider
the average minimum projected separation between clusters as a function of cluster
mass. We denote this quantity, corrected for galaxy inclination, by ${\bar D}_{\rm
min}(M)$, where the {\it bar} denotes the average for all clusters of a particular
mass $M$, and the subscript ``min'' denotes the minimum distance $D$ to these other
cluster. For a random distribution of cluster positions, this separation is about
equal to the inverse square root of the average projected cluster density. In
addition, we denote the number of clusters in a logarithmic mass interval by
$N(M)d\log M$. For a uniform random distribution of clusters of all masses, ${\bar
D}_{\rm min}(M)$ multiplied by $N(M)^{1/2}$ is independent of $M$. The mass
distribution function for clusters and incompleteness at low mass both enter
${\bar D}_{\rm min}(M)$ and $N(M)$ in the same way, cancelling out.

If high mass clusters of mass $M_{\rm H}$ are more bunched together than low mass
clusters of mass $M_{\rm L}$, then
\begin{equation}
N(M_{\rm H})^{1/2}{\bar D}_{\rm min}(M_{\rm H})<N(M_{\rm L})^{1/2}{\bar D}_{\rm min}(M_{\rm L}).
\end{equation}
Thus, a plot of $N(M)^{1/2}{\bar D}_{\rm min}(M)$ versus $M$ indicates
the relative segregation of different masses.

For comparisons among different regions in a galaxy or different galaxies, the above
quantity should be normalized to the region size, which we take to be the average
projected separation (corrected for galaxy inclination) between the clusters in the
lowest mass interval, ${\bar D}(M_{\rm low})$.  This interval is chosen because
generally it has the most clusters and gives the most accurate region size. Thus,
the quantity to consider as a function of mass is the {\it relative average minimum
projected separation} (RAMPS) corrected for galaxy inclination,
\begin{equation}
RAMPS(M) =N(M)^{1/2}{\bar D}_{\rm min}(M)/{\bar D}(M_{\rm low}).
\end{equation}

Another measure of relative cluster separation is the two-point correlation function
\citep[e.g.,][]{bastian09,grasha17b}, which determines the relative proportions of
all separations. The RAMPS differs because it uses only the nearest distances.

\subsection{Testing the RAMPS}

A fractal hierarchical model shows the trends in RAMPS. This model is made on a
$512\times512$ square grid of total size 1 with 8 levels of hierarchy starting with
a $4\times4$ grid of cells at the top in level $i=2$ \citep[see][]{elmegreen18}. At
each level $i$ there is a probability $p$ of choosing a cell that will be further
subdivided into $2\times2$ cells in the next lower level. This probability depends
on the fractal dimension $D_{\rm f}$ and is given by $p=2^{D_{\rm f}-2}$. To choose
a cell, a random number between 0 and 1 is generated and compared to $p$; if the
random number is smaller than $p$, we choose the cell. Note that a fractal dimension
$D_{\rm f}=2$ causes all cells to be chosen ($p=1$), filling the square grid
completely in two-dimensions. For the model we use $D_{\rm f}=1.3$ because that
matches the observations of interstellar clouds \citep{elmegreen96}. Each level
sub-divides only the cells chosen at the next higher level. For level $i$ from 2 to
9, the size of the cell is $1/2^i$ and the total number of cells is $2^i \times2^i$,
although only the fraction $p$ are chosen at each level.

We assign a mass to each of the cells at all levels, considering that the cell
center represents the position of a star cluster in the hierarchy of young stellar
structures. To give mass segregation, uniformity, or inverse segregation, we let the
cluster mass $M$ scale with the level $i$ such that $M=M_0\mu^i$ where
$\mu=10^{0.5}$, 1, or $10^{-0.5}$ for $i=2,...,6$ in these three cases respectively.
To keep the masses in the range of $10^3\;M_\odot$ to $10^5\;M_\odot$, we set
$M_0=10^2\;M_\odot$ and $10^6\;M_\odot$ for the segregation and inverse segregation
cases. For the uniform case, we let $\log M_0$ be a random number uniformly
distributed between 3 and 5. Because cells at levels with larger $i$ are on average
closer together within their hierarchies than cells at lower $i$ within the same
hierarchy, the segregation of high mass clusters to denser average regions
corresponds to a greater proportion of massive clusters at large $i$. This is why
$\mu>1$ corresponds to mass segregation and $\mu<1$ corresponds to inverse mass
segregation, where massive clusters systematically avoid each other compared to low
mass clusters.


Figure 1 shows the RAMPS function for the three values of $\mu$. Mass segregation
with $\mu>1$ has a negative slope and inverse mass segregation has a positive slope.
Random mass in the hierarchy corresponds to a horizontal line in the figure. The way
to interpret the negative slope is that the nearest neighbor high mass cluster to
another high mass cluster is closer than the average cluster spacing would be at
that mass if they were randomly distributed over the whole region.


\section{Data}
\label{data}

Catalogs in the Hubble Space Telescope LEGUS survey were used to obtain the
positions, ages and masses of measured clusters \citep{calzetti15,adamo17},
considering Padova stellar evolution models with starburst extinction curves. To
keep the sample as free as possible from fading effects with age, we
consider only clusters more massive than a distance-dependent limit, $M_{\rm
limit}$, and younger than 125 Myr. For almost all galaxies in LEGUS, the lower limit
to the detectable mass at 125 Myr age is $94D_{\rm Mpc}^2\;M_\odot$ for distance
$D_{\rm Mpc}$ in Mpc. This mass corresponds to an absolute V-band magnitude of
$M_{\rm v}=-6.0$. For a typical distance of 6 Mpc, the typical mass limit is
$3400\;M_\odot$. Then, for the entire sample, there are 14 galaxies that have 35 or
more such clusters in classes 2 or 3 and which span a factor of $10^{1.5}$ or more
in mass. These galaxies and their cluster counts are listed in Table 1 along with
the RAMPS slopes. Galaxy distances, star formation rates and stellar masses are from
\cite{calzetti15}, as are inclinations (not listed). Galaxy position angles are from
various sources such as the Third Reference Catalog of Bright Galaxies
\citep{devauc91} and the LITTLE THINGS survey \citep{hunter12} when available, and
measured from LEGUS images at the LEGUS web
site\footnote{\url{https://legus.stsci.edu/legus_observations.html}} when not
available, all verified by measurements on the Digitized Sky
Survey\footnote{\url{http://archive.stsci.edu/cgi-bin/dss_form}}.

Table 1 also lists the 11 galaxies that have 35 or more class 1 clusters spanning a
factor of $10^{1.5}$ or more in mass within the same age and mass limits. Similarly,
the table lists the cluster counts and RAMPS slopes considering only class 2 types
alone and only class 3 types alone. Class 1 clusters are compact, class 2 are
somewhat elongated, and class 3 are multi-core \citep{adamo17}. The galaxies span a
factor of $\sim100$ in star formation rate and stellar mass, so they represent a
fair sample of spiral and dwarf galaxy types.

The clusters were first divided into logarithmic mass intervals of 0.5 dex;
the number of bins in Table 1 represents the number of these intervals spanned by
the cluster masses. As for the random trial discussed above, the projected
separations between each cluster and all the other clusters in the same mass
interval (corrected for galaxy inclination) were determined and the minimum of these
separations was noted. This minimum represents the distance between each particular
cluster and its nearest cluster of the same mass. The average of these minimum
distances was then determined for each mass interval. After multiplication by the
square root of the number of clusters in the mass interval and division by the
average separation in the lowest mass interval, we obtain the RAMPS.

Figure 2 shows the RAMPS for the LEGUS clusters in classes 2 and 3, divided into
those with increasing or nearly constant average slopes on the left and those with
decreasing average slopes on the right. These slopes are determined from the whole
mass range, which, e.g., for 4 bins, corresponds to a factor of $\sim100$.

The middle panels in Figure 2 show the RAMPS slope versus the star formation rate
(SFR) (left) and the galaxy stellar mass $M_{\rm star}$ (right).  There is an
increasing trend in both panels.  The slopes of these trends were determined by
least squares linear fits with uncertainties given by the student-t distribution at
90\% probability. Corresponding $\chi^2$ values are from the sum of the squared
differences between the RAMPS slopes and the linear fits, normalized to the
uncertainties in the slopes. The result is a slope of $0.14\pm0.13$ with $\chi_{\rm
r}^2=6.5$ for 12 degrees of freedom (DOF) in the plot of RAMPS slope versus
$\log(SFR)$, and $0.14\pm0.10$ with $\chi^2=4.9$ for 12 DOF in the plot of RAMPS
slope versus $\log M_{\rm star}$. These $\chi^2$ values are smaller than the number
of DOF, which is number of RAMPS values minus 2, indicating reasonably good
fits. A summary is in Table 1.

The bottom panels in Figure 2 show the RAMPS slopes calculated with an equal
number of clusters in each of 6 cluster mass intervals (rather than equal $\log(M)$
intervals). The upward trends are present for this binning too, although slightly
smaller. Still, their slopes (Table 1) are within the error bars of the slopes in the
middle panels.


The correlations between the slope of the RAMPS and the galaxy SFR or stellar mass
imply that low mass galaxies have their most massive class 2-3 clusters closer
together than average, whereas high mass galaxies have all cluster masses randomly
distributed.

Inclination errors could affect the results, but a recalculation of the class
2+3 case with zero inclinations gave about the same slopes: $ 0.16\pm 0.14$ versus
$\log{SFR}$ and $0.16\pm 0.11$ versus $\log(M_{\rm star})$. The $\chi^2$ values were
much higher without inclination corrections, however, 31 and 18, respectively.

The clusters were identified by eye for all galaxies, but for NGC 5194, clusters
were also identified with Machine Learning (ML) techniques, using the visual
identifications as a training set \citep{messa18,grasha19}. There are more ML
clusters than visual clusters for this galaxy, but the slope of the RAMPS is about
the same in both cases. The green points in the bottom of Figure 2 and the green
crosses and line in the top left are for clusters identified by ML in NGC 5194,
compared to the blue points at the same SFR and stellar mass on the bottom and the
blue crosses on the top left.

We also considered clusters in the tidal dwarf galaxies connected with the
interacting galaxy NGC 5291. These were obtained from the study by \cite{fensch19}
with distances between the clusters determined by assuming zero inclination and
position angle as these are highly irregular galaxies. Among their sample of 272
clusters with masses less than $3\times10^5\;M_\odot$ and not category 0 (which are
excluded because they have more than two HST passbands with only upper limits on the
flux), we include all 106 clusters with masses larger than $10^4\;M_\odot$ and ages
less than 100 Myr. These limits avoid the loss of clusters from fading. There are
several dwarfs in the collision debris but we can treat all of them as one large
distribution to derive the slope of the RAMPS because the nearest cluster to any
given cluster of the same mass is likely to be inside the same tidal dwarf. The
other factors in RAMPS, $N(M)$ and $D_{\rm low}$, are constant and do not affect the
slope. The red segmented line and red points in Figure 2 show the results for NGC
5291. The RAMPS for NGC 5291 has a negative slope like the other dwarf galaxies.

Figure 3 shows $M_{\rm star}$ versus distance, with the expected trend for lower
mass galaxies, which are more common, to be closer. Because of this, the closer
galaxies also have more negative slopes in the RAMPS, as shown in the right-hand
panel.  These distance trends are not the cause of the varying slopes for RAMPS,
however. The cluster separations are well resolved for all of the galaxy distances,
and the cluster mass ranges are about the same scaled for distance. Moreover, the
distance to NGC 5291 (red points in Figure 3) is larger than even the massive
galaxies in LEGUS and yet the slope of its RAMPS is negative, like the closer
dwarfs.

The RAMPS for class 1 clusters in LEGUS is shown in Figure 4. Again the positive and
negative slopes are separated in the two top panels, but now the slopes are all
around 0, as also shown in the middle and bottom panels. Evidently the class 1
clusters have different grouping properties than the class 2 and 3 clusters.
Separate plots for class 2 clusters and for class 3 clusters alone (not shown)
repeat the correlation in Figure 2, which was for the combined classes. The number
of clusters and RAMPS slopes for these separate classes are given in Table 1 for
completeness.

We also determined the RAMPS for only the young class 1 clusters with ages less than
30 Myr (not shown); these had no obvious trends with galaxy stellar mass either,
although there were only 6 galaxies with enough young clusters to plot.  The
slope of the RAMPS slope versus log(SFR) linear fit for young class 1 clusters is $
0.11\pm 0.28$ ($\chi^2=0.8$, DOF$=4$), but the range in SFR is only a factor of 7.
Versus $\log(M_{\rm star})$, the slope is $ 0.05\pm 0.18$ ($\chi^2=0.9$, DOF$=4$)
with a factor of 28 range in $M_{\rm star}$.

We also checked whether class 1 and class $2+3$ clusters show a significant
correlation between RAMPS and the SFR per unit area.  The areas were determined
from the distances and from $D_{25}$ in \cite{devauc91} or the NASA/IPAC
Extragalactic Database. No trends were found. There is a correlation with area
alone, however, similar to that in Figures 2 and 4, such that more massive clusters
are more clumped together in smaller galaxies, which are also the lower mass
galaxies in the previous figures.

Table 1 summarizes for all cluster classes the slopes of the linear fits,
$S\pm\epsilon$, the $\chi^2$ values, and the number of DOF for the RAMPS slopes
versus $\log(SFR)$, $\log(M_{\rm star})$, $\log(Area)$, and $\log(SFR/Area)$.

\section{Discussion}
\label{discussion}

The RAMPS method suggests that class 2 and 3 star clusters in LEGUS, which are
somewhat elongated or irregular in shape, are mass segregated in low-mass galaxies,
whereas class 1 clusters, which are compact, are not. Clusters in the tidal dwarfs
around NGC 5291 are also mass segregated.

Dwarf galaxies differ from high-mass galaxies in ways that could account for the
trends. For example, dwarfs have weaker gravitational potentials than massive
galaxies, so the mutual attraction between massive clouds and the clusters they form
is larger in proportion to background tidal forces for a dwarf galaxy. The tidal
dwarfs around NGC 5291 could also be devoid of dark matter, which makes their
background gravitational potential even weaker. This implies that massive clouds and
clusters can move closer to each other, or accrete more interstellar gas, with less
of an influence from galactic shear in lower mass galaxies or tidal dwarfs. This
explanation does not obviously account for the lack of mass segregation by class 1
clusters, however.

Alternatively, high mass clusters could scatter away their low mass neighbors more
effectively when Coriolis forces are low, leaving the high mass clusters more
concentrated in each star-forming region than average. This explanation could
include the observed difference between cluster classes because class 1 clusters
might scatter better than class 2-3 clusters, which could break apart during the
process. Class 1 clusters in LEGUS are also older on average than class 2 and 3
clusters \citep{grasha17b}, giving them more time to scatter. However, the young
class 1 clusters, less than 30 Myr, did not show a correlation with galaxy stellar
mass. Class 1 clusters could be a random selection of clusters that are dense.

Other models for cluster mass segregation in small galaxies could be developed
around a possibly larger Jeans mass for fragments near the center of a star-forming
gas complex, or a higher gas density in higher-mass clouds. Numerical simulations
need to address these possibilities.

\acknowledgments The authors are grateful to the referee for comments on the
manuscript. Based on observations made with the NASA/ESA Hubble Space Telescope,
obtained at the Space Telescope Science Institute, which is operated by the
Association of Universities for Research in Astronomy, Inc., under NASA contract NAS
5-26555. These observations are associated with programs \#13364 (LEGUS) and \#14727
(NGC 5291), including grant HST-GO-14727.004-A to BGE. This research has made use of
the NASA/IPAC Extragalactic Database (NED), which is funded by the National
Aeronautics and Space Administration and operated by the California Institute of
Technology.

\newgeometry{margin=0.35in}
\begin{landscape}
\begin{deluxetable}{lcclcccccccccccc}
\tabletypesize{\scriptsize} \tablecolumns{10} \tablewidth{0pt}
\tablecaption{Cluster Sample}
\tablehead{
\colhead{Galaxy$^1$}&
\colhead{D$^2$}&
\colhead{SFR}&
\colhead{$M_{\rm star}$}&
&
\colhead{1$^3$}&
&
&
\colhead{2}&
&
&
\colhead{3}&
&
&
\colhead{$2+3$}&
\\
&
&
\colhead{}&
&
\colhead{N$^4$}&
\colhead{B}&
\colhead{S}&
\colhead{N}&
\colhead{B}&
\colhead{S}&
\colhead{N}&
\colhead{B}&
\colhead{S}&
\colhead{N}&
\colhead{B}&
\colhead{S}
}
\startdata
   628 &   9.90 &   3.67 &  11  &  133 &  4 & \tiny$   0.02\pm  0.15$  &   59 &  4 & \tiny$   0.05\pm  0.28$  &   44 &  3 & \tiny$  -0.15\pm  0.53$ &  103 &  4 & \tiny$   0.07\pm  0.37$\\
  1249 &   6.90 &   0.15 &  0.55&    - &  - &       -  &    - &  - &       -  &    - &  - &       - &   43 &  4 &   \tiny$   0.28\pm  0.35$\\
  1313 &   4.39 &   1.15 &  2.6 &  113 &  4 & \tiny$  -0.10\pm  0.13$  &  250 &  4 & \tiny$   0.01\pm  0.13$  &  286 &  4 &   \tiny$   0.06\pm  0.16$ &  536 &  4 & \tiny$  -0.01\pm  0.06$\\
  1566 &  13.20 &   5.67 &  27  &  212 &  4 & \tiny$   0.02\pm  0.26$  &  119 &  3 & \tiny$   0.14\pm  0.63$  &   93 &  3 &   \tiny$   0.33\pm  0.47$ &  212 &  3 & \tiny$   0.09\pm  0.47$\\
  1705 &   5.10 &   0.11 &  0.13&    - &  - &       -  &    - &  - &       -  &    - &  - &       - &   36 &  4 & \tiny$  -0.18\pm  0.92$\\
  3344 &   7.00 &   0.86 &  5.0 &   35 &  3 & \tiny$   0.22\pm  0.30$  &    - &  - &       -  &    - &  - &       - &    - &  - &       -\\
  3351 &  10.00 &   1.57 &  21  &   39 &  5 & \tiny$  -0.64\pm  0.18$  &    - &  - &       -  &    - &  - &       - &    - &  - &       -\\
  3627 &  10.10 &   4.89 &  31  &  144 &  3 & \tiny$   0.11\pm  0.03$  &   92 &  4 & \tiny$   0.06\pm  0.11$  &   50 &  4 &   \tiny$   0.23\pm  0.35$ &  142 &  4 & \tiny$   0.19\pm  0.32$\\
  3738 &   4.90 &   0.07 &  0.24&    - &  - &       -  &   60 &  4 & \tiny$  -0.43\pm  0.23$  &   55 &  3 & \tiny$  -0.13\pm  0.89$ &  115 &  4 &  \tiny$  -0.36\pm  0.23$\\
  4395 &   4.30 &   0.34 &  0.60&    - &  - &       -  &    - &  - &       -  &    - &  - &       - &   44 &  4 & \tiny$  -0.04\pm  0.35$\\
  4449 &   4.31 &   0.94 &  1.1 &   66 &  5 & \tiny$   0.04\pm  0.31$  &  160 &  5 & \tiny$  -0.39\pm  0.13$  &  132 &  4 & \tiny$  -0.07\pm  0.20$ &  292 &  5 &  \tiny$  -0.38\pm  0.21$\\
  4656 &   5.50 &   0.50 &  0.40&   43 &  3 & \tiny$  -0.61\pm  0.17$  &   61 &  3 & \tiny$  -0.31\pm  1.17$  &   46 &  4 & \tiny$   0.27\pm  0.20$ &  107 &  5 &  \tiny$  -0.03\pm  0.31$\\
  5194 &   7.66 &   6.88 &  24  &  140 &  5 & \tiny$  -0.06\pm  0.38$  &  198 &  5 & \tiny$   0.09\pm  0.11$  &   81 &  3 & \tiny$   0.51\pm  0.95$ &  279 &  5 &   \tiny$   0.14\pm  0.09$\\
  5194-ML & 7.66&   6.88 &  24  &  610 &  5 & \tiny$  -0.26\pm  0.29$  &  646 &  4 & \tiny$  -0.11\pm  0.16$  &   89 &  3 & \tiny$   0.36\pm  0.62$ &  735 &  4  & \tiny$  -0.08\pm  0.17$\\
  5253 &   3.15 &   0.10 &  0.22&    - &  - &       -  &    - &  - &       -  &   67 &  5 &  \tiny$  -0.23\pm  0.45$ &   93 &  5 & \tiny$  -0.18\pm  0.21$\\
  6503 &   5.27 &   0.32 &  1.9 &   40 &  3 & \tiny$   0.06\pm  0.20$  &   37 &  3 & \tiny$-0.19\pm  0.05$  &   59 &  3 & \tiny$   0.04\pm  0.09$ &   96 &  3 &   \tiny$   0.02\pm  0.13$\\
  7793 &   3.44 &   0.52 &  3.2 &   41 &  3 & \tiny$  -0.18\pm  0.13$  &  116 &  3 & \tiny$  -0.11\pm  0.44$  &   85 &  4 & \tiny$   0.02\pm  0.30$ &  201 &  4 &   \tiny$   0.09\pm  0.24$\\
  5291$^5$&63.5 &   0.15 &  2.00&  106 &  6 &  -0.65  &&&&&&&&&\\

  \hline
  SFR:$^6$ &\tiny$S\pm \epsilon$&$\chi^2$&\tiny DOF&\tiny $ 0.14\pm 0.35$ & 37 & 9 & \tiny$ 0.28\pm 0.11$ & 7.2 & 8 & \tiny$ 0.22\pm 0.15$ & 3.3 & 9 & \tiny$ 0.14\pm 0.13$ &  6.5 &  12\\
  Mass: &\tiny$S\pm \epsilon$&$\chi^2$&\tiny DOF&\tiny $ 0.10\pm 0.26$ & 37 & 9 & \tiny$ 0.25\pm 0.07$ & 3.3 & 8 & \tiny$ 0.17\pm 0.14$ & 3.6 & 9 & \tiny$ 0.14\pm 0.10$ &  4.9 &  12\\
  Area: &\tiny$S\pm \epsilon$&$\chi^2$&\tiny DOF&\tiny $-0.10\pm 0.38$ & 79 & 9 & \tiny$ 0.25\pm 0.14$ & 3.8 & 8 & \tiny$ 0.24\pm 0.15$ & 0.9 & 9 & \tiny$ 0.20\pm 0.10$ &  4.1 &  12\\
  \tiny SFR/A:&\tiny$S\pm \epsilon$&$\chi^2$&\tiny DOF&\tiny $ 0.45\pm 0.42$ & 34 & 9 & \tiny$ 0.12\pm 0.39$ & 15  & 8 & \tiny$-0.02\pm 0.42$ & 3.0 & 9 & \tiny$-0.18\pm 0.22$ &  14  &  12\\
  \hline
  SFR:$^7$ &\tiny$S\pm \epsilon$&$\chi^2$&\tiny DOF&\tiny $ 0.06\pm 0.39$ & 14 & 9 & \tiny$ 0.10\pm 0.08$ &  2.7 & 8 & \tiny$ 0.09\pm 0.10$ & 1.6 & 9 & \tiny$ 0.08\pm 0.07$ &  5.4 &  12\\
  Mass: &\tiny$S\pm \epsilon$&$\chi^2$&\tiny DOF&\tiny $ 0.05\pm 0.29$ & 14 & 9 & \tiny$ 0.10\pm 0.06$ &  1.9 & 8 & \tiny$ 0.06\pm 0.09$ & 1.6 & 9 & \tiny$ 0.07\pm 0.06$ &  4.3 &  12\\
\enddata
\tablenotetext{1}{NGC numbers except for 1249, which is a UGC number. NGC 5194 has a
first row for clusters identified by eye and a second row ("ML") for cluster
identified by machine learning.} \tablenotetext{2}{Distance in Mpc, star formation
rate (SFR) in $M_\odot$ yr$^{-1}$, Galaxy stellar mass $M_{\rm star}$ in
$10^9\;M_\odot$.} \tablenotetext{3}{Cluster classes: 1 is compact, class 2 is
elongated and class 3 is multi-core}. \tablenotetext{4}{N represents the number of
clusters in this class with a mass exceeding a certain distance-dependent limit and
age less than 126 Myr (log age = 8.1); B represents the number of mass bins for the
clusters, each of width 0.5 dex; S is the slope of the RAMPS, i.e., the derivative
of the log of the RAMPS with respect to log of the cluster mass for equal intervals
of the log of the cluster mass.} \tablenotetext{5} {NGC 5291 is the system of tidal
dwarfs, which have a total of 106 usable clusters that have not been classified
according to the LEGUS system.} \tablenotetext{6} {These four rows give the slopes
and errors ($S\pm\epsilon$) of the RAMPS slopes versus $\log(SFR)$, $\log(M_{\rm
star})$, $\log(Area)$ (in ${\rm kpc}^2$), and $\log(SFR/Area)$ (in $M_\odot$
kpc$^{-2}$ yr$^{-1}$), where area is measured at the radius of 25 magnitudes per
square arcsec in the B band. These rows also show the $\chi^2$ values and the number
of degrees of freedom, $DOF$, which is the number of RAMPS values minus 2.}
\tablenotetext{7} {These next two rows are for RAMPS with equal numbers of clusters
per interval of the cluster mass.} \label{galaxylist}
\end{deluxetable}
\end{landscape}
\restoregeometry

\begin{figure}
\epsscale{5.}
\includegraphics[width=4.in]{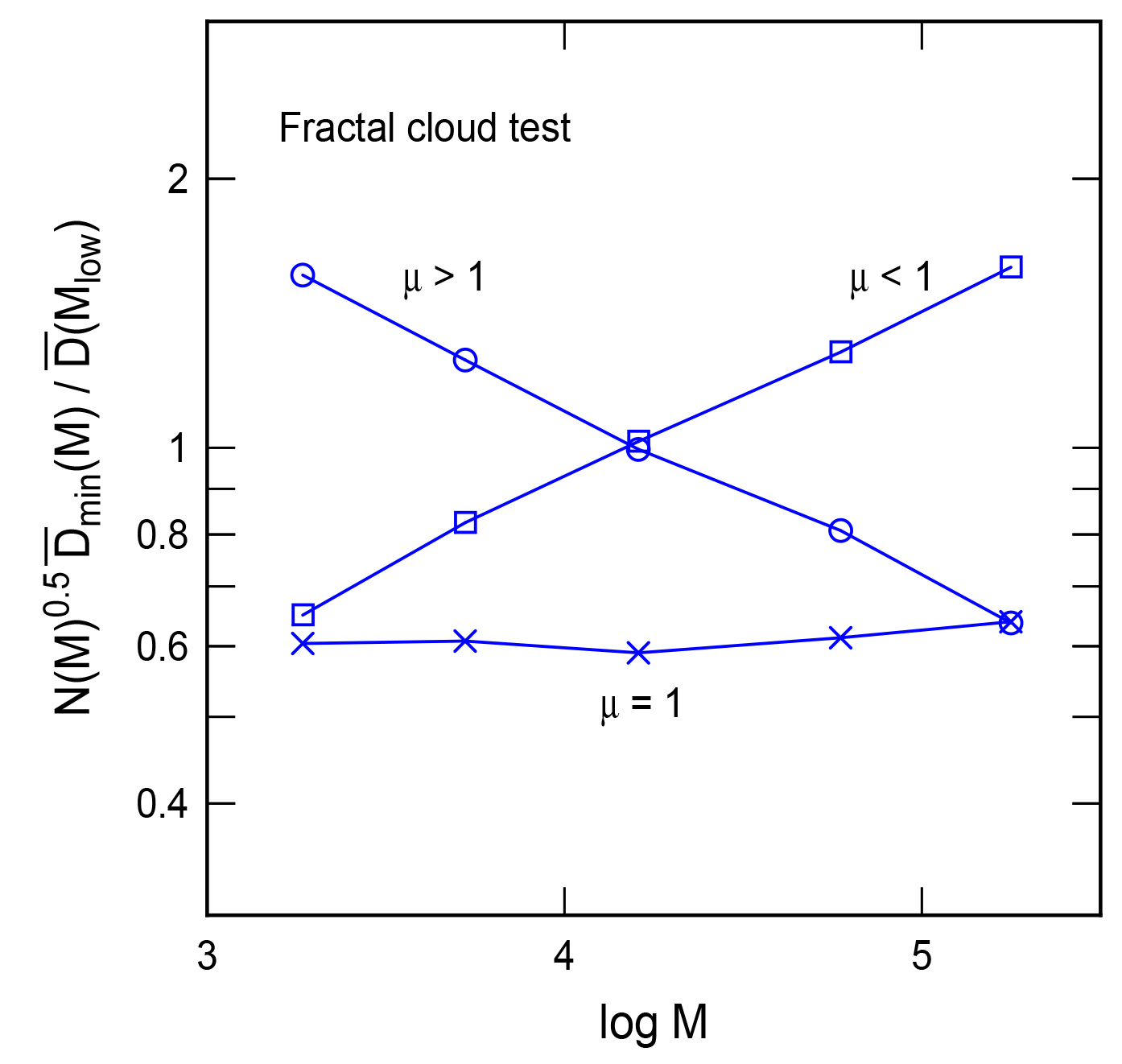}
\caption{(top) The relative average minimum projected separations (RAMPS)
between clusters are plotted
versus the cluster mass for randomly positioned clusters in a hierarchical
fractal distribution.  The mass units are arbitrary.
Increasing slopes indicate that low-mass clusters
are preferentially clumped together, while decreasing slopes indicate that high-mass
clusters are clumped together. Randomly positioned clusters give a flat slope.}
\label{fensch_mass_correl_9999}
\end{figure}

\begin{figure}
\epsscale{5.}
\includegraphics[width=4.in]{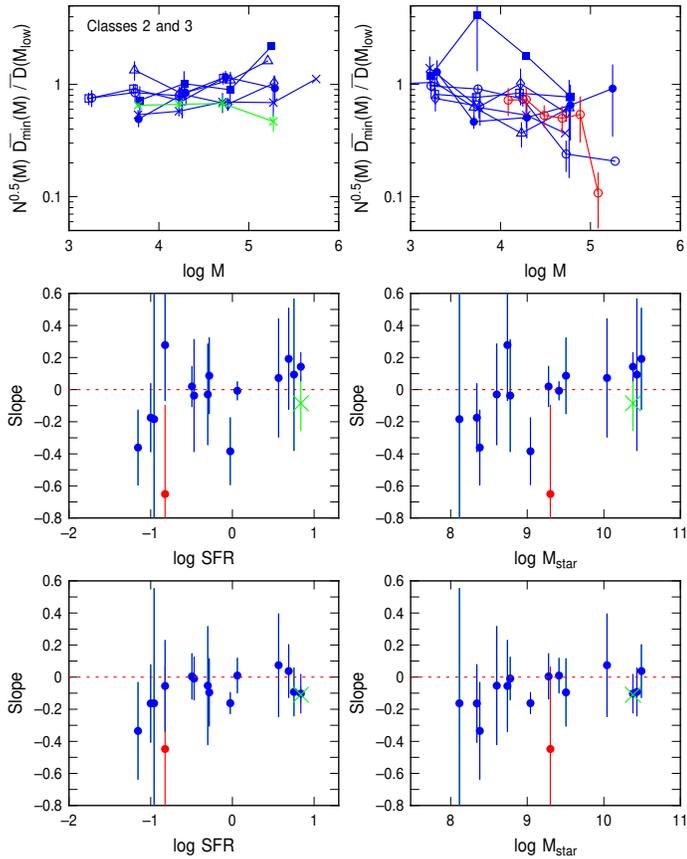}
\caption{(top)
The RAMPS are plotted
versus the cluster mass $M$ (in $M_\odot$) for LEGUS clusters in classes 2 and 3
(blue lines) and for
NGC 5291 (red line) clusters. The RAMPS are dimensionless.
Error bars are from the student-t distribution
at 90\% probability; they are symmetric around the linear value of the RAMPS but
asymmetric here when plotted logarithmically (points with no
error bars have only 2 clusters in that mass interval).
Cluster masses are randomly offset from the center of the
mass bin for clarity; different galaxies have different symbols. Increasing RAMPS are
on the left and decreasing RAMPS are on the right.  (middle, bottom) The slopes of the
RAMPS are shown versus the total galaxy star formation rates (in
$M_\odot$ yr$^{-1}$) and galaxy stellar masses (in $M_\odot$). The middle panels are for RAMPS
calculated using
equal $\log(M)$ intervals, as in the top panels,  while the bottom panels are for RAMPS
calculated using equal numbers of clusters
in each mass interval.
Low-mass galaxies tend to have decreasing RAMPS, indicating a greater tendency for
high mass clusters to collect together compared to low mass clusters. The green line and crosses
on the top left and green crosses in the middle and bottom panels are for NGC 5194 clusters determined by Machine
Learning; the blue crosses on the top left are for NGC 5194 visual identifications.}
\label{fensch_mass_correl_legus_class2-3_ML}
\end{figure}

\begin{figure}
\epsscale{5.}
\includegraphics[width=6.in]{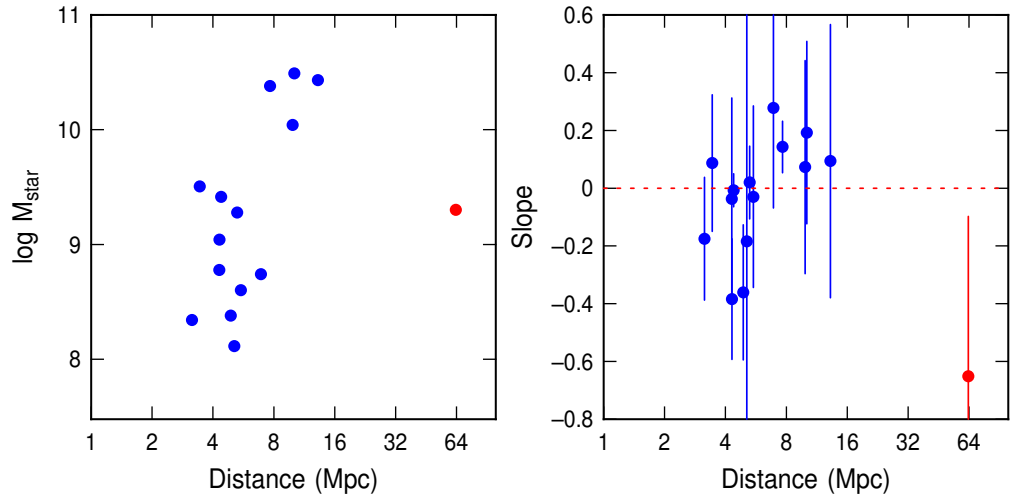}
\caption{(left) Galaxy stellar mass in $M_\odot$ is shown as a function of
galaxy distance in Mpc. (right) The slope of the RAMPS for class 2 and 3 clusters
is shown as a function of distance.
The blue points are LEGUS galaxies and the red point is for tidal dwarfs around
NGC 5291. For the LEGUS galaxies,
there is a correlation between galaxy mass and distance, and so a resulting
correlation appears between RAMPS slope and distance, but the RAMPS does not appear to be
biased for distance by itself. }
\label{fensch_distance}
\end{figure}

\begin{figure}
\epsscale{5.}
\includegraphics[width=5.in]{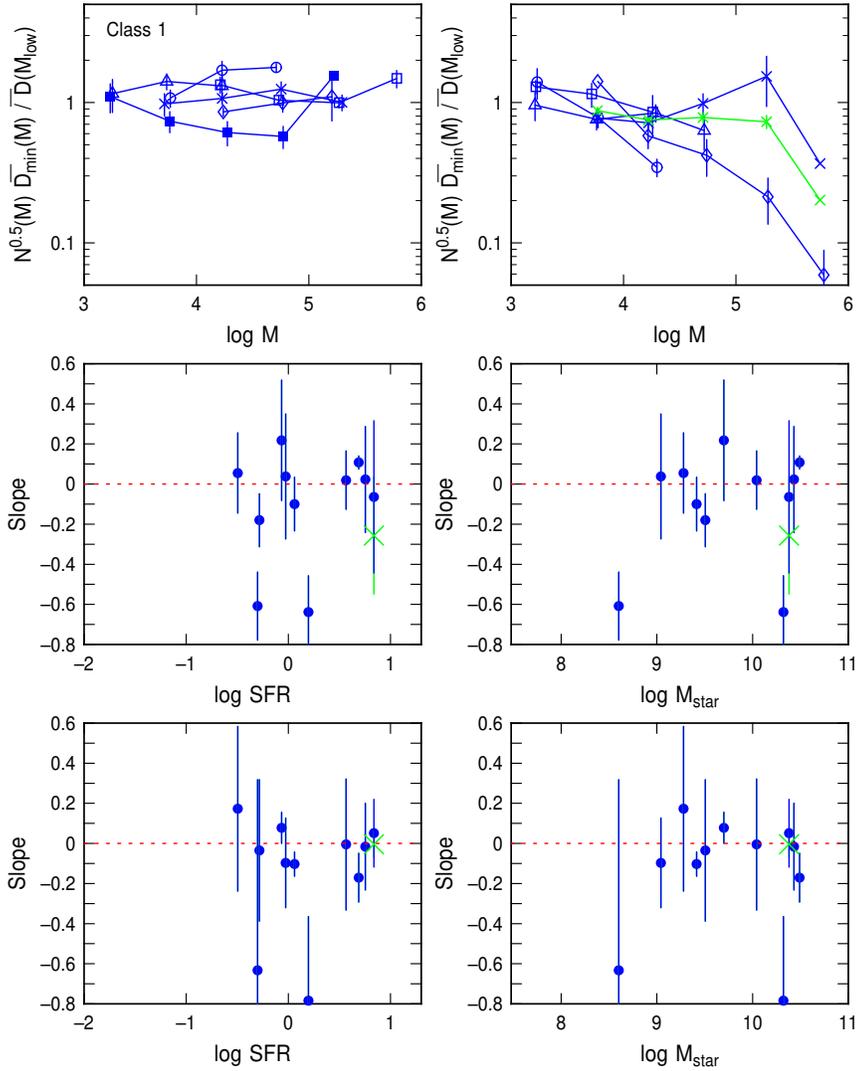}
\caption{(top) The RAMPS are plotted
versus the cluster mass ($M_{\odot}$) for LEGUS clusters in class 1. (middle, bottom) The slopes of the
RAMPS are shown versus the total galaxy star formation rates (in $M_\odot$ yr$^{-1}$)
and galaxy stellar masses (in $M_\odot$).
Unlike the class 2 and 3 clusters in Figure 2, the class 1 clusters
appear to be randomly distributed for all galaxy masses. The green line and crosses
on the top right and the green crosses in the middle and bottom panels
are for NGC 5194 clusters discovered by
Machine Learning. As in Figure 2, the middle panels are for RAMPS calculated with equal
$\log(M)$ intervals of cluster mass, and the bottom panels are for RAMPS calculated with
equal numbers of clusters in each mass interval.}
\label{fensch_mass_correl_legus_class1_ML}
\end{figure}

\end{document}